\documentclass[preprint]{aastex}

\slugcomment{Accepted to {\it The Astrophysical Journal}}

\shorttitle{GRB 000301C}
\shortauthors{Rhoads \& Fruchter}

\begin{document}

\def\erg{\:\hbox{erg}}
\def\cm{\:\hbox{cm}}
\def\sec{\:\hbox{s}}
\def\day{\:\hbox{day}}
\def\days{\:\hbox{days}}
\def\micron{\:\mu m}
\def\mm{\:\hbox{mm}}
\def\cm{\:\hbox{cm}}
\def\mJy{\:\hbox{mJy}}
\def\microJy{\:\mu \hbox{Jy}}
\def\mag{\:\hbox{mag}}
\def\GHz{\:\hbox{GHz}}
\def\Hz{\:\hbox{Hz}}
\def\keV{\:\hbox{keV}}
\newcommand\basi{BASI}
\def\dof{\:\hbox{d.o.f}}

\title{The Near Infrared and Multiwavelength Afterglow of GRB 000301c}


\author{James E. Rhoads \and Andrew S. Fruchter}
\affil{ Space Telescope Science Institute}
% , 3700 San Martin Drive, Baltimore, MD 21218 ; rhoads@stsci.edu}

\begin{abstract}
We present near-infrared observations of the counterpart of
GRB 000301c.  The K' filter ($2.1 \micron$) light curve shows a
well-sampled break in the decay slope at $t \approx 3.5$ days
post-burst.  The early time slope is very shallow ($\sim -0.1$), while
the late time slope is steep ($-2.2$).
Comparison with the optical (R band) light curve shows marginally significant
differences, especially in the early time decay slope (which is steeper
in the optical) and the break time (which occurs {\it later} in the
optical). 
This is contrary to the general expectation that light curve
breaks should either be achromatic (e.g., for breaks due to collimation
effects) or should occur later at longer wavelengths (for most other breaks).
The observed color variations might be intrinsic to the afterglow,
or might indicate systematic errors of $\ga 0.08$ magnitude in all fluxes.
Even if the break is achromatic, we argue that its sharpness poses
difficulties for explanations that depend on collimated ejecta.
The R light curve shows further signs of fairly rapid variability (a
bump, steep drop, and plateau) that are not apparent in the K' light
curve.
In addition, by combining the IR-optical-UV data with millimeter and
radio fluxes, we are able to constrain the locations of the
self-absorption break and cooling break and to infer the location of
the spectral peak at $t = 3$ days, $f_\nu \approx 3.4\mJy$ at
$\nu \approx 10^{12} \Hz$.
Using the multiwavelength spectral energy distribution, we are able to
constrain the blast wave energy, which was $E \ga 3 \times 10^{53}
\erg$ if the explosion was isotropic.  This implies a maximum gamma
ray production efficiency of $\sim 0.15$ for GRB 000301C.
\end{abstract}


\keywords{Gamma rays--- bursts}

\section{Introduction}
Infrared observations can be used to improve our understanding of
gamma ray burst afterglows in several ways.  First, they can be
combined with optical measurements to obtain spectral slope
measurements with a much wider wavelength baseline, and hence yield a more
accurate spectral slope than optical data alone.  Second, they can be used
to test observed
light curve breaks for wavelength dependence, which is an important
discriminant between breaks due to ejecta collimation and other
possible causes.  Finally, it has been suggested
that bursts are preferentially located in dusty regions,
e.g., under ``hypernova'' scenarios where bursters are a final
evolutionary stage of some class of massive stars.
If so, and if the dusty region extends beyond the expected dust
destruction distance (Waxman \& Draine 1999),
then near-infrared observations will detect some afterglows that are
obscured at optical wavelengths.  At less extreme dust optical depths,
near-infrared data help to characterize the host galaxy extinction and so infer
both extinction corrected fluxes and properties of dust and gas
in high redshift GRB host galaxies.

We present here near-infrared (NIR) photometric observations of the
afterglow of the gamma ray burst GRB 000301c.  These constitute the
best-sampled near infrared light curve for any afterglow to date.

GRB 000301c was detected independently by the All-Sky Monitor on the
Rossi X-Ray Timing Explorer and by two spacecraft (Ulysses and NEAR)
of the current Interplanetary Network on 2000 March 1.4108 UT.  The
event was a single peaked GRB lasting approximately 10 seconds (Smith,
Hurley, \& Cline 2000) at low energies ($< 10 \keV$) and 2 seconds at
higher energies ($> 25 \keV$) (Jensen et al 2000).  Coordinates were
available approximately 36 hours after the burst, and an optical
counterpart was reported by Fynbo et al (2000) based on observations
at March 3.21 UT.  The redshift of the burst was first reported as $z
= 1.95 \pm 0.1$ (Smette et al 2000a) and subsequently refined to $z =
2.028 \pm 0.025$ (Smette et al 2000b) using the observed Lyman break
in a near-UV spectrum of the GRB afterglow.  Weak metal lines in
absorption yielded further improvements to $z = 2.0335 \pm
0.0003$ (Castro et al 2000) or $z=2.0404 \pm 0.0008$ (Jensen et al
2000).

\section{Photometric Data}
\section{Infrared Telescope Facility Images}
We present data obtained at the NASA Infrared Telescope Facility
(IRTF) on Mauna Kea as part of a service mode Target of Opportunity
program for broadband near-IR followup of gamma ray bursts using the
NSFCam imager.  Observing conditions were good throughout the period,
with clear skies and subarcsecond seeing on all nights.  All data were
taken with a plate scale of $0.3''$ per pixel.  A log of the
observations is given in table~\ref{obslog}.

\begin{table}
\begin{tabular}{lcccccccc}
UT Date & $t$ & Filter & Exposure & Number & Seeing &
 Magnitude & Error & Error \\
 & & & (seconds) & of Frames & & & (photometric) & (total) \\
% 4.652 & 3.241 & J  & 700  & 14  & $0.95''$  & $19.10 \pm 0.05$ \\
% Correcting the foregoing for color transformation to CIT system from
% the NSFCam web pages, we have
4.652 & 3.241 & J  & 700  & 14  & $0.95''$  & $19.11$ & $0.05$ & $0.064$\\
4.640 & 3.229 & K' & 672  & 14  & $0.85''$  & $17.66$ & $0.04$ & $0.057$\\
5.610 & 4.199 & K' & 672  & 14  & $0.69''$  & $18.01$ & $0.05$ & $0.064$\\
6.595 & 5.184 & K' & 435  &  9  & $0.71''$  & $18.57$ & $0.12$ & $0.126$\\
8.590 & 7.179 & K' & 1400 & 28  & $0.68''$  & $19.29$ & $0.09$ & $0.098$
\end{tabular}
\caption{Log of the IRTF data.  $t$ is the observed time elapsed since the
GRB.  ``Exposure'' is the sum of the exposure times for all GRB images
in that filter and night. The ``photometric'' error is due to
poisson statistics of received photons.  The ``total'' error combines
the photometric error in quadrature with an estimated $0.04$
magnitudes error due to sky subtraction and flatfielding
uncertainties.
\label{obslog}}
\end{table}

We reduced and analyzed the data following standard near-IR
procedures.  Raw sky flats were generated for each filter and night
using the mean of all available frames after outlier rejection to
eliminate the influence of objects on the flatfield.  Final flats were
generated by subtracting stacked dark frames (taken with the same
exposure time, number of coadds, and number of nondestructive reads as
the data) from the raw flats and normalizing the result to have mean
$=1$.  Individual frames were then sky-subtracted (using the median of
three to six frames taken immediately before and/or after the object
frame) and flatfielded.  Frames were aligned using bilinear
interpolation to implement a simple shift based on the measured
centroid of a bright star.  Finally, the aligned frames were combined
using a clipped mean after removing any sky subtraction residuals
through subtraction of the modal pixel value.

We measured the GRB fluxes on all nights using aperture photometry
with an aperture diameter of $1.8''$.  All afterglow photometry was
taken relative to the bright star $5.7''$ west and $1''$ south of the
optical transient.  (This is star A of Garnavich et al 2000a.)  The J
and K' band magnitudes of this star were $J_{\hbox{CIT}}=16.64 \pm
0.01$ and $K'=15.97 \pm 0.02$.  These were calibrated using
observations of UKIRT faint standard 27 (FS27, Casali \& Hawarden
1992) taken immediately before the first IRTF observations of GRB
000301c.  The reference magnitudes used for FS 27 were $K'=13.14$
(based on photometric transformation equations from Wainscoat \& Cowie
1992) and $J_{\hbox{CIT}} = 13.45$ (based on equations from the NSFCam
documentation).
The standard star observations were processed in the same way as the
GRB images, and a larger photometry aperture ($5.4''$ diameter) was
used for flux calibration, to reduce sensitivity to any centering
errors or seeing variations between the standard star and GRB frames.
No correction was made for atmospheric extinction; however, any such
correction would be small ($\la 0.01 \mag$), because the standard was
observed at airmass 1.09, and the GRB field at airmass 1.02.  The
reference star showed no evidence for variability either in our
near-IR data or in optical data from two groups (Garnavich, private
communication; Halpern, private communication).  The aperture
photometry included a local sky estimation and subtraction using the
mode of pixel values in an annulus around each point source.  This
step should control any residual sky level or first order gradient in
the sky.  Moreover, by selecting the annulus to avoid bright objects
and using the mode for sky level estimation, we also control the
possible influence of other sources on sky level estimates.

Our counts always remained below the nominal linearity
limit for NSFCam (Leggett \& Denault 1996); we therefore did not apply
nonlinearity corrections to our data.  We placed the J band data on
the CIT magnitude system using $J_{\hbox{CIT}} = J_{\hbox{MK}} - 0.01
(J-K)$,
% and on the UKIRT magnitude system using $J_{\hbox{UKIRT}} =
% J_{\hbox{MK}} + 0.073 (J-K)$,
where $J_{\hbox{MK}}$ is the natural
magnitude system for the current Mauna Kea J filter in NSFCam.
This gives $J_{\hbox{CIT}} = 19.11$ for the afterglow on March 4.652.
The K' magnitudes were left on the NSFCam instrumental system.

Errors due to photon counting statistics were computed based on an
iteratively clipped variance of each night's final stacked images,
suitably corrected for the correlated noise introduced by bilinear
interpolation.  Sky subtraction errors arise only from the difference
between sky level in the photometry aperture and in the sky annulus.
These errors are separated into statistical errors (due to photon
noise in the sky annulus) and systematic (due to objects in the sky
annulus or any other source of bias in estimating the true background
under the transient source).  The statistical part is always small
compared to statistical errors from the object flux measurement, due
to the large number of pixels in the sky annulus ($\sim 900$) and much
smaller number ($\sim 28$) in the photometric aperture.  Systematic
errors in sky subtraction are potentially larger, but we believe they
are under reasonable control in our data set because the sky level was
removed in two steps (globally, during data reduction, and locally,
during aperture photometry) accounting for up to first order gradients
in both time and space; and because the weather was cooperative, with
relatively little temporal variation in sky background ($<10\%$
minimum to maximum in any night and filter, and usually less) and
subarcsecond seeing (allowing small photometry apertures).  Further
possible error sources include centroiding of the afterglow and
reference star (expected to be systematic-error limited at the $0.1$
pixel level) and residual flatfielding difficulties.  Neither will be
large compared to our photon counting noise.
To be conservative, we estimate that sky subtraction and flatfielding
errors combined may affect all of our photometry at up to the $4\%$ level.
Table~\ref{obslog} lists both pure photon counting errors and error
bars including this systematic error added in quadrature.

% Centering errors in the photometry may also affect the photometry.
% The centroid location of a point source can be estimated with accuracy
% of order $\theta_\gamma / sqrt(n_\gamma)$, where $\theta_\gamma \la
% 0.4''$ is the dispersion in photon location and $n_\gamma \ga 3\times
% 10^4$ is the number of photons detected.  This gives 0.002 arcsec,
% or 0.007 pixels...   So really we're systematic limited here to
% 0.1 pixel or so.

\subsection{Data from Other Sources}
In addition to the IRTF data, we include in our analysis data from
other observatories presented in the literature (both the GRB
Coordinate Network Circulars and preprints by Masetti et al 2000 and
Jensen et al 2000).  These data are summarized in table~\ref{litlog}.
For data points reported multiple times, we use the most recently
reported value.  In particular, for the Uttar Pradesh State
Observatory data we take values from Masetti et al (2000) rather than
Sagar et al (2000).

% Justify uniformity of other K' data
Two early time K' data points come from Calar Alto data of
Stecklum et al (2000) and Subaru data of Kobayashi et al (2000a,b).  Flux
values from both were measured relative to the Garnavich et al (2000a)
star A.  The use of a uniform reference star allows the data from
multiple observatories to be compared with reasonable confidence.
Residual differences in color terms should be small since all three
observatories used the same photometric bandpass.  To allow for color
terms, we have used $\sigma = 0.03 \mag$ as the effective error on the
Subaru data, rather than the photometric error $0.01 \mag$ reported by
Kobayashi et al (2000a,b).
%
% K' band color terms:  
% Calar Alto Omega Cass camera/spectrograph uses a 1024^2 HgCdTe HAWAII array.
% Subaru IRCS uses an Aladdin II 1024^2 InSb array.
% IRTF NSFCam uses a 256^2 InSb array.
% Can't find a quantitative indication of the QE(lambda) for the
% different arrays, but as the reference stars are all in the
% Rayleigh-Jeans limit in K band, we compare nu^-2 with nu^-1 and the
% likely color terms are comparable to the O star case for R band
% photometry, i.e. <~ 0.05 mag.

% Justify uniformity of optical data

Similarly, the optical data reported in the literature and used in
this paper has all been calibrated to either stars A-D of Garnavich et
al (2000a) (for some R band data) or to the photometry of Henden et al
(2000) (for other optical filters and the rest of the R band data).
The R band fluxes measured by Henden et al for stars A-D agree with
those from Garnavich et al within the uncertainties, which are $\sim
5\%$.  The mean and median magnitude differences between the two
calibrations are $0.050$ and $0.036$ magnitude respectively, in the
sense that Garnavich et al report brighter magnitudes than do Henden
et al.  The possibility of inconsistent photometric zero points in
different data sets therefore amounts to about $0.05$ magnitude
between authors using calibrations from the two alternative sources,
and less for authors using the same calibration.  The two largest sets
of uniformly reduced optical data currently available are Masetti et
al (2000) and Jensen et al (2000).  Both groups use the Henden et al
calibration.  We have increased all R band fluxes reported on the
Garnavich et al (2000a) zero point by $0.04$ magnitude to adjust all R
band photometry to the Henden et al (2000) zero point.

Another possible systematic difference between photometry from
different observatories is color term differences.  To estimate the
importance of these effects, we compare the magnitude difference
between the GRB afterglow and a comparison star for thick and thinned
CCD chips behind a standard Cousins R filter (Bessell 1990).  The
quantum efficiency of a thick chip was approximated by $\hbox{QE} =
0.45 + (\lambda / \micron - 0.65)$, and that of a thin chip by
$\hbox{QE} = 0.85 - 0.8 (\lambda / \micron - 0.65)$.  (Both these
relations are approximations based on plots from the Steward
Observatory CCD Lab web pages and valid for the R band spectral
region.) We take the
afterglow spectrum to be a power law with $f_\nu \propto \nu^{-1}$ (a
characteristic value for this afterglow before any reddening
corrections; see below), and take stellar spectral energy
distributions from Gunn and Stryker (1983).  The result depends on the
color of the comparison star.  For stars with the colors of unreddened
G0 to K4 dwarfs (which includes stars A and D of Garnavich et al
2000a), the difference in inferred magnitude for the two
$\hbox{QE}(\lambda)$ models is $\le 0.01$ magnitude.  For redder or
bluer stars, the differences can be substantial: Up to $0.05$
magnitude for O stars, $0.07$ magnitude for M0 stars (approximately
matching Garnavich et al's stars B and C), and $0.20$ magnitude for
the coolest M stars.  However, only $9\%$ of stars in the Henden et
al list are redder than stars B and C.  Overall, then, color terms will be
a modest source of error (no greater than the random photometric error
and the zero point error) except for pathological choices of
comparison stars.  Jensen et al (2000) explicitly address this issue
by using comparison stars of color similar to the GRB afterglow in
their analysis.

% http://www.noao.edu/kpno/manuals/dim/node23.html  0.03 (V-R) is color
%  term for R filter at KPNO 2.1m with T1KA thinned AR coated backside
%  treated CCD.
% http://www.physics.nau.edu/~pmassey/Mosaicphot.html  for color terms
%  for the CCD Mosaic camera (though only thinned chips, and only
%  UBV filters) showing how much effect different chips can have.
%  It's about 0.03 (B-V) for the max difference between two chips in V band.
% The Steward CCD Lab http://sauron.as.arizona.edu/ has plots of QE(lambda)
%  for both thick and thin chips.  These give a behaviors ranging from
%  front illuminated chips with
%     QE = 0.45 + 1.0*(lambda-0.65) (lambda in microns)
%  to QE = 0.85 - 0.8*(lambda-0.65) for a thinned chip
%   Both approximations valid over 0.6 to 0.7 microns but let's extrapolate
%   to the whole R band, the result will be more conservative than if we
%   used the real behavior outside 0.6 to 0.7 microns.
%

A final possible source of error arises from possible ``crowded
field'' effects in the photometry.  Star A is located $6''$ from the
GRB afterglow.  With sufficiently poor seeing, aperture photometry of
the afterglow might be affected by the wings of this star's point
spread function (PSF).  We have estimated the magnitude of this error as
a function of aperture size and seeing through numerical calculations with
a Gaussian PSF model.  We find that photometric errors from this
source are negligible ($\ll 0.01 \mag)$ for seeing better than $2''$ (FWHM).
They remain $\le 0.015 \mag$ even for $3''$ seeing provided that the
photometric aperture radius used is $\le 1.8''$.
PSF-fitting photometry, as done by Masetti et al (2000) and Jensen et
al (2000), is more robust to such errors.

% Discuss the UV data.
The ultraviolet flux at $0.28 \micron \le \lambda \le 0.33 \micron$ is
taken from the continuum level in the Hubble Space Telescope STIS
spectrum of the afterglow (Smette et al 2000b).  This measurement was
$f_\lambda = (7.3^{+0.8}_{-1.8}{\scriptsize \pm 0.6}) \times 10^{-18}
\erg \cm^{-2} \sec^{-1} \mbox{\AA}^{-1}$.  The dominant (first) error bar
comes from uncertainties in wavelength calibration coupled with
strongly wavelength-dependent sensitivity; the second error bar is a
random error.  This corresponds to $f_\nu(0.305 \micron) =
(2.26^{+0.25}_{-0.56}{\scriptsize \pm 0.19}) \microJy$.  Over this
wavelength range, $f_\lambda$ is approximately constant, so $f_\nu
\sim \nu^{-2}$.  This slope is steeper than the optical-IR slopes
derived in section~\ref{sedsec} below, but is consistent with a
wavelength independent intrinsic slope if intergalactic Lyman line
absorption is considered (Madau 1995).

\begin{deluxetable}{lllllll}
% Note:  T_0 = 1.4108
% \begin{tabular}{lllllll}
\tablecaption{Data from the literature.
\label{litlog}}
\tablehead{
\colhead{UT Date} & \colhead{$t$} & \colhead{Filter} & 
\colhead{Magnitude} & \colhead{Authors} & \colhead{Reference}}
\startdata
\hline
% Near IR data -- K' filter
3.215 & 1.804 & K' & $17.52 \pm 0.06$ & Stecklum et al &
   GCNC 572 \& private communication \\
3.55  & 2.14  & K' & $17.53 \pm 0.03$ & Kobayashi et al & GCNC 577, 587 \\
4.640 & 3.229 & K' & $17.66 \pm 0.057$ & Rhoads \& Fruchter & This work \\
5.610 & 4.199 & K' & $18.01 \pm 0.064$ & Rhoads \& Fruchter & This work \\
6.595 & 5.184 & K' & $18.57 \pm 0.126$ & Rhoads \& Fruchter & This work \\
8.590 & 7.179 & K' & $19.29 \pm 0.098$ & Rhoads \& Fruchter & This work \\
\hline
% Near IR data -- J filter
3.55  & 2.14  & J  & $18.91 \pm 0.03$ & Kobayashi et al & GCNC 577, 587 \\
4.652 & 3.241 & J  & $19.11 \pm 0.064$ &Rhoads \& Fruchter & This work \\
\hline
% Optical data -- R band
2.906 & 1.495 & R & $20.42 \pm 0.04$ & Masetti et al & A \\
2.961 & 1.550 & R & $20.02 \pm 0.03$ & Bhargavi \& Cowsik & GCNC 630 \\
3.14  & 1.729 & R & $20.09 \pm 0.04$ & Jensen et al & B \\
3.144 & 1.733 & R & $20.25 \pm 0.05$ & Masetti et al & A \\
% 3.17  & 1.76  & R & $19.94 \pm 0.04$ & Fynbo et al & GCNC 570, 576 \\
3.17  & 1.759 & R & $20.15 \pm 0.04$ & Jensen et al & B \\
% 3.191 & 1.780 & R & $20.11 \pm 0.05$ & Bernabei et al & GCNC 599 \\
3.185 & 1.774 & R & $20.16 \pm 0.05$ & Masetti et al & A \\
3.19  & 1.779 & R & $20.11 \pm 0.04$ & Jensen et al & B \\
3.205 & 1.794 & R & $20.25 \pm 0.05$ & Masetti et al & A\\
3.21  & 1.799 & R & $20.14 \pm 0.04$ & Jensen et al & B \\
3.25  & 1.839 & R & $20.16 \pm 0.04$ & Jensen et al & B \\

3.51  & 2.10  & R & $20.32^\dag \pm 0.05$ & Garnavich et al & GCNC 573 \\
% 3.51  & 2.10  & R & $20.24 \pm 0.07$ & Veillet et al & GCNC 575 \\
3.51  & 2.10  & R & $20.27 \pm 0.04$ & Veillet et al & GCNC 575, 588 \\
3.51  & 2.10  & R & $20.28^\dag \pm 0.05$ & Halpern et al & GCNC 578 \\
% 3.93  & 2.52  & R & $20.53 \pm 0.05$ & Mohan et al & GCNC 595 \\
3.913 & 2.502 & R & $20.51 \pm 0.04$ & Masetti et al & A \\
3.998 & 2.587 & R & $20.49 \pm 0.10$ & Bhargavi \& Cowsik & GCNC 630 \\
% 4.057 & 2.646 & R & $20.46 \pm 0.09$ & Castro-Tirado et al & GCNC 579 \\
4.038 & 2.627 & R & $20.53 \pm 0.06$ & Masetti et al & A \\
4.080 & 2.669 & R & $20.613^\dag \pm 0.06$ & Gal-Yam et al & GCNC 593 \\
% 4.178 & 2.767 & R & $20.22 \pm 0.2$  & Bernabei et al & GCNC 599 \\
4.38  & 2.97  & R & $20.60^\dag \pm 0.05$ & Garnavich et al & GCNC 581 \\
4.42  & 3.01  & R & $20.61 \pm 0.06$ & Jensen et al & B \\
4.458 & 3.047 & R & $20.58^\dag \pm 0.06$ & Mujica et al & GCNC 597 \\
4.48  & 3.069 & R & $20.58 \pm 0.03$ & Jensen et al & B \\
4.49  & 3.079 & R & $20.54 \pm 0.04$ & Jensen et al & B \\
4.50  & 3.089 & R & $20.60 \pm 0.04$ & Jensen et al & B \\
4.50  & 3.09  & R & $20.65^\dag \pm 0.04$ & Halpern et al & GCNC 582 \\
4.909 & 3.498 & R & $20.58 \pm 0.05$ & Bhargavi and Cowsik & GCNC 630 \\
5.135 & 3.724 & R & $20.47 \pm 0.07$ & Masetti et al & A \\
5.39  & 3.979 & R & $20.61 \pm 0.05$ & Jensen et al & B \\
5.63  & 4.22  & R & $20.86 \pm 0.04$ & Veillet et al & GCNC 588 \\
% 5.96  & 4.55  & R & $21.18 \pm 0.05$ & Mohan et al & GCNC 595 \\
5.930 & 4.519 & R & $21.14 \pm 0.06$ & Masetti et al & A \\
% 6.145 & 4.734 & R & $21.60 \pm 0.2$  & Bernabei et al & GCNC 599 \\
6.135 & 4.724 & R & $21.65 \pm 0.20$ & Masetti et al & A \\
6.22 &4.81 &``R'' & $21.5  \pm 0.15$ & Fruchter et al & GCNC 602 \\ %STIS
6.39 &4.98 &``R'' & $21.43 \pm 0.26$ & Jensen et al & B \\
% 7.135 & 5.724 & R & $21.63 \pm 0.15$ & Bernabei et al & GCNC 599 \\
7.125 & 5.714 & R & $21.68 \pm 0.15$ & Masetti et al & A \\
7.22  & 5.809 & R & $21.59 \pm 0.07$ & Jensen et al & B \\
7.65  & 6.24  & R & $21.70 \pm 0.07$ & Veillet et al & GCNC 598 \\
7.894 & 6.483 & R & $22.00 \pm 0.15$ & Masetti et al & A \\
% 8.157 & 6.746 & R & $21.63 \pm 0.1$  & Bernabei et al & GCNC 599 \\
8.146 & 6.735 & R & $21.68 \pm 0.10$ &  Masetti et al & A \\
8.18  & 6.769 & R & $21.80 \pm 0.05$ &  Jensen et al & B \\
8.924 & 7.513 & R & $22.04 \pm 0.20$ &  Masetti et al & A \\
9.15  & 7.739 & R & $22.11 \pm 0.15$ &  Jensen et al & B \\
9.52  & 8.11  & R & $22.32^\dag \pm 0.09$ & Halpern \& Kemp & GCNC 604 \\
11.39 & 9.979 & R & $23.12 \pm 0.18$ &  Jensen et al & B \\
11.63 & 10.22 & R & $23.02 \pm 0.10$ & Veillet et al & GCNC 610  \\
12.44 & 11.03 & R & $23.10 \pm 0.22$ & Jensen et al & B \\
14.60 & 13.19 & R & $23.82 \pm 0.10$ & Veillet et al & GCNC 611 \\
34.9  & 33.5  &``R''&$26.5 \pm 0.15$ & Fruchter et al& GCNC 627, 701 \\
50.0  & 48.59 &``R''&$27.9 \pm 0.15$ & Fruchter et al & GCNC 701 \\
% Optical data -- V band
% See GCNC 583 for zero pointing information from Henden et al.
% See GCNC 593 for a V band point (Gal-Yam et al)
% Optical data -- B band
\hline
3.50  & 2.09  & B & $21.11 \pm 0.04$ & Veillet et al & GCNC 575,588 \\
4.52  & 3.11  & B & $21.41 \pm 0.04$ & Halpern et al & GCNC 585 \\
14.60 & 13.19 & B & $24.83 \pm 0.12$ & Veillet et al & GCNC 611 \\
% UV data
\hline
6.714 & 5.303 & UV & $2.26^{+0.31}_{-0.59} \microJy$ & Smette et al & C \\
\hline
% Radio data--  Berger, Frail, et al give 0.3 mJy at March 5.67 UT at 8.46 GHz
% 4.385 & 2.974 & 1.2mm & $1.9 \pm 0.3 \mJy$ & Bertoldi & GCNC 580 \\
4.385 & 2.974 & 1.2mm & $2.1 \pm 0.3 \mJy$ & Berger et al & D \\
\hline
% 5.67 & 4.26 & 3.54cm & $315 \pm 16 \microJy$ & Berger \& Frail & 
%    GCNC 589 \& private communication
5.67 & 4.26 & 1.33cm & $884 \pm 216\microJy$ & Berger et al & D \\
5.67 & 4.26 & 3.54cm & $316 \pm 41 \microJy$ & Berger et al & D \\
5.67 & 4.26 & 6.17cm & $240 \pm 53 \microJy$ & Berger et al & D \\
5.67 & 4.26 & 20.9cm &  $11 \pm 79 \microJy$ & Berger et al & D 
\enddata
% \end{tabular}
\tablecomments{ {\scriptsize The estimated error on Kobayashi
et al data has been increased from their $0.01$ magnitude statistical
error to $0.03$ mag to allow for possible color terms in conversion
between their photometric system and ours.  R band magnitudes flagged
with ``$^\dag$'' were originally reported with calibration from
Garnavich et al (GCNC 573) and are here adjusted by $+0.04 \mag$ to
place them on the flux zero point of Henden et al (GCNC 583).  Four
measurements use ``pseudo-R'' filters: Three of these (from Fruchter
et al) are HST STIS unfiltered data, and one (from Jensen et al) is
from VLT spectrophotometry; both are calibrated to R band.  References
are either GRB Coordinate Network Circular numbers, or preprints coded
as follows: A: Masetti et al 2000, to appear in \aap,
astro-ph/0004186v2; B: Jensen et al 2000, submitted to \aap,
astro-ph/0005609; C: Smette et al 2000, submitted to \apj,
astro-ph/0007202; D: Berger et al, to appear in \apj,
astro-ph/0005465.  } }
\end{deluxetable}

\section{Light curve fitting}
\begin{figure}
\plotone{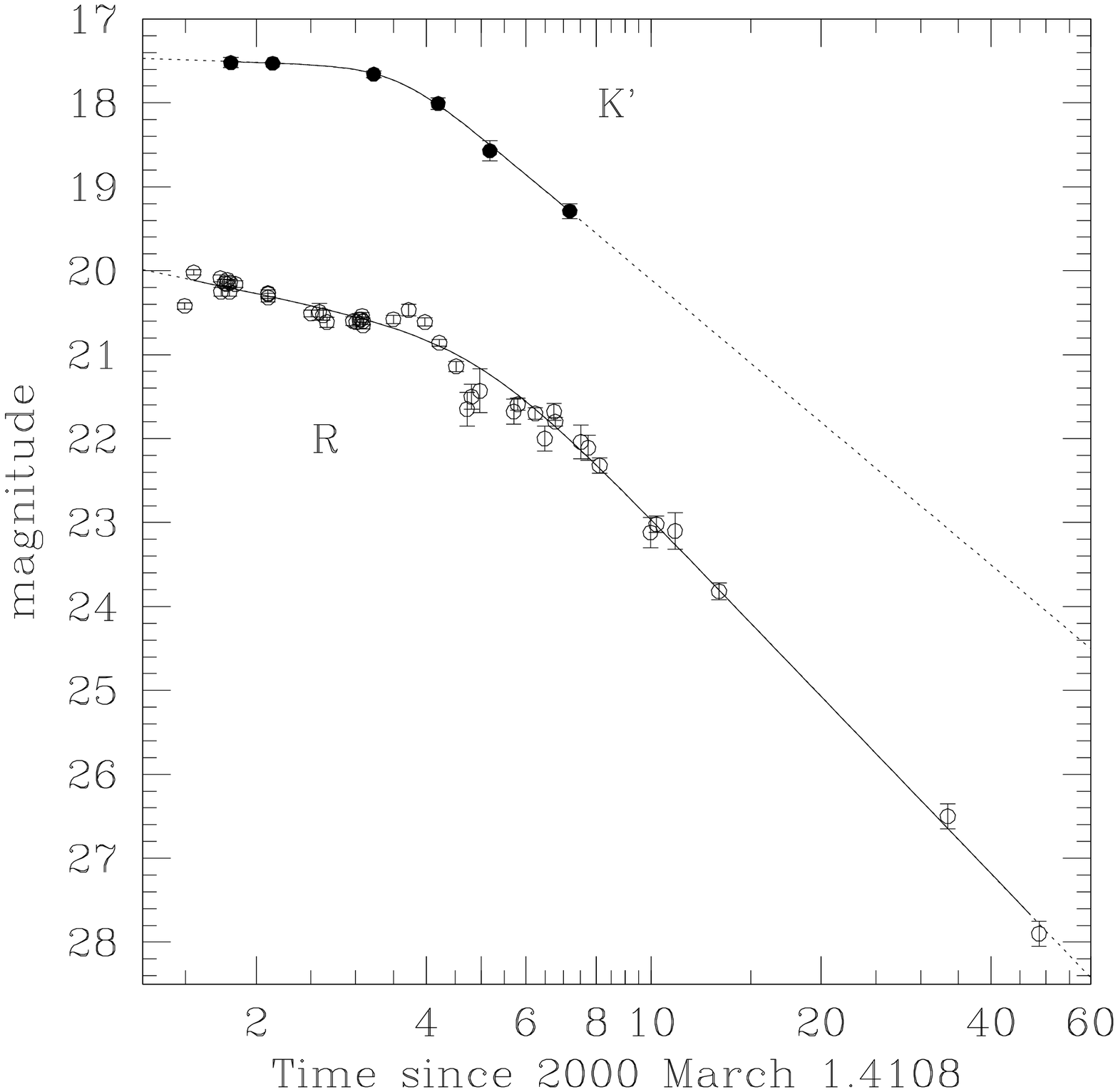}
\caption{
The light curves of the GRB 000301c afterglow in both K' and R
filters.  The data are summarized in table~\ref{litlog}.  Smoothly
broken power law fits using the empirical fitting form in
equation~\ref{sbpl} are shown as solid lines where they interpolate
the data, and as dotted lines where they are extrapolations.  The
fitted parameters are given in the text.  Filled circles are K' filter
data and open circles are R filter data.
\label{lcfig}}
\end{figure}

Both the K' and R light curves are shown in figure~\ref{lcfig}.
The K' band light curve shows a smooth rollover from an initially constant
flux to a rapid decay at late time.  The R band light curve shows a
qualitatively similar behavior, but the early time R flux decays
more steeply than the early K' flux, and there are additional
irregularities in the R band light curve that are arguably
significant (e.g., Masetti et al 2000; Sagar et al 2000).

Broken power laws can be empirically fitted by functions of the form
$f = f_0 \left[ (t/t_b)^{\alpha_1 \beta} + (t/t_b)^{\alpha_2 \beta}
\right]^{-1/\beta}$.  With $\alpha_1 < \alpha_2$ and $\beta > 0$, this
function describes a light curve falling as $t^{-\alpha_1}$ at $t\ll
t_b$ and $t^{-\alpha_2}$ at $t\gg t_b$.  $\beta$ controls the
sharpness of the break, with larger $\beta$ implying a sharper
break. The function that Stanek et al (1999) used to fit the light
curve of GRB 990510 is the special case $\beta=1$ of this function.
Rewritten in magnitudes, the fitting function becomes
\begin{equation}
m = m_b + {2.5 \over \beta} \left\{ \log_{10}\left[ (t/t_b)^{\alpha_1
\beta} + (t / t_b)^{\alpha_2 \beta} \right] - \log_{10}(2) \right\}
\label{sbpl}
\end{equation}
where $m_b$ is the magnitude at time $t_b$.

The K' light curve can be well fitted by this model.  The best fit
values (with $\chi^2 = 0.422$ for one degree of freedom [$\dof$]) are
$t_b = 3.57$ days, $K_b = 17.76$, $\alpha_1 = 0.09$, $\alpha_2 =
2.26$, and $\beta = 4.23$.

The R band data show significant departures from such a light curve.
The first R band data point is clearly fainter than the extrapolation
of the other early R data.  If we exclude this point, the best fit to
the R data has $\chi^2 / \dof = 2.8$ for 41 degrees of freedom.  This
fit has $t_b = 5.1$ days, $R_b = 21.18$, $\alpha_1 = 0.69$, $\alpha_2
= 2.77$, and $\beta = 3.2$.  (Including the earliest point gives
instead $\chi^2 / \dof \approx 4.7$, with $t_b = 4.97$ days, $R_b =
21.15$, $\alpha_1 = 0.56$, $\alpha_2 = 2.81$, and $\beta = 2.25$.)
The largest deviations from the smoothe light curve (aside from the
first data point) are from the bump at $t \approx 3.75$ days, the
plateau at $\la t \la 7$ days, and the steep decline between the two.
We believe all three are likely real features.  This behavior may
indicate variations in the external medium density (see Kumar \&
Panaitescu 2000b) or refreshed shock effects (Panaitescu, Meszaros, \&
Rees 1998; Dai \& Lu 2000).  Either explanation suggests that the K'
data should show similar effects.  To test this, we examine the
behavior of $R-K'$ below.

If we assume a fixed $R-K'$ color and compare the $K'$ data to the R
band light curve model, the best fit color is $R-K' = 2.80$ and
$\chi^2 / \dof = 6.6$ for 5 degrees of freedom (6 data points - 1
color).  The same time interval includes about half the R data (31
points), and the data $-$ model residuals in R yield $\chi^2 / \dof =
3.9$.  This relatively large $\chi^2$ is due to the presence of the
bump, drop, and plateau in this time interval.  In short, the smooth R
band fit does not describe the R data very well in this time period,
but does an even worse job describing the K' data.
% Note, used 27.6 effective degrees of freedom (out of 41 total)
% for 31 R data points (out of 46 total).

\section{$R - K'$ color variations} \label{colsec}
To assess the significance of the light curve fit differences above,
we have examined in detail the $R-K'$ colors of the afterglow for the
six $K'$ data points.  At each epoch, we use selected R data points to
obtain the most reliable interpolated or extrapolated R flux at the
time of the K' measurement.  All measurements are adjusted to the zero
point of Henden et al (2000).  Details for each epoch follow.

March 3.215: We interpolate between the March 3.21 and March 3.25
Nordic Optical Telescope measurements by Jensen et al (2000),
obtaining $R=20.145 \pm 0.04$.  Hence, $R-K' = 2.62 \pm 0.07$.

March 3.55: We take a weighted average of the three measurements from
March 3.51 (Garnavich et al 2000a; Halpern et al 2000a; Veillet et al
2000a,b).  We apply corrections to place the first two epochs on the
Henden et al zero point, and further add $0.02$ magnitudes to adjust
the R flux to the $K'$ epoch, one hour later.  (This corresponds to a
local $t^{-1}$ decay.)  We obtain $R=20.32 \pm 0.04$ and $R-K' = 2.79
\pm 0.05$.

March 4.64: We combine four R measurements from March 4.48 to 4.50
(three from Jensen et al 2000; one from Halpern et al 2000b, adjusted
by 0.05 mag to the Henden et al zero point) and interpolate to the
March 4.909 point of Bhargavi \& Cowsik to obtain $R=20.59 \pm 0.04$.
Hence, $R-K' = 2.93 \pm 0.06$.

March 5.61:  We extrapolate the March 5.63 R band point from Veillet
et al (2000) to the K' epoch, obtaining $R = 20.85 \pm 0.04$ and $R-K'
= 2.84 \pm 0.06$.

March 6.595: This epoch is during the ``plateau'' in the R band light
curve (Bernabei et al 2000).  Fitting a single power law through the R
data from March 6.135 to March 7.22, evaluating it at March 6.595, and
estimating the flux error through Monte Carlo simulations, we find
$R=21.57 \pm 0.07$. The resulting color is $R-K' = 3.00 \pm 0.14$.
This color does depend on our assessment that the ``plateau'' is a
real feature.  If we extend the fit to earlier times (starting at
March 5.93), the color becomes $0.14$ mag bluer, though a single power
law fit does not describe the data well over the full period from
March 5.93 to March 7.22.

March 8.590: We use the R data from
March 8.146 through 9.52.  We again  fit a single power law decay
through all the R data, evaluate it at the epoch of the K'
measurement, and determine the error bar from simulations.  The result
is $R = 21.95 \pm 0.04$, so that $R-K' = 2.66 \pm 0.10$.

\begin{figure}
\plotone{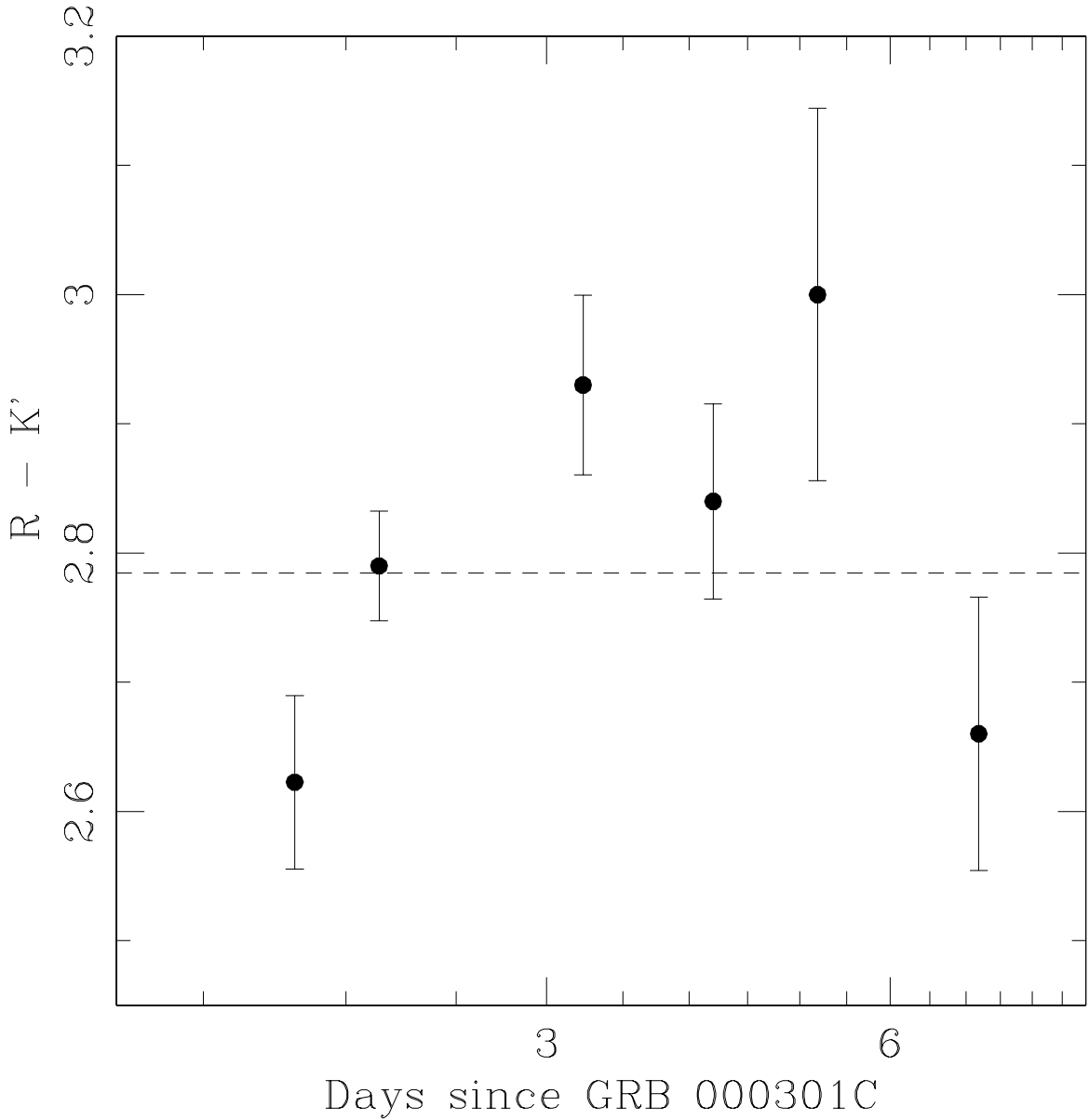}
\caption{
The $R-K'$ color evolution of the GRB 000301c afterglow.  The dashed
line shows the best fit single color, $R-K' = 2.78$.  Each epoch is a
direct $K'$ measurement and an interpolated $R$ flux.
\label{rkfig}}
\end{figure}

These values of $R-K'$ are plotted as a function of time in
figure~\ref{rkfig}.  If we assume a single $R-K'$ color throughout
the afterglow, the best fit is $R-K' = 2.79$.  With $\chi^2 / \dof = 2.87$
for $5$ degrees of freedom, this fit is not especially good.

We consider three possible explanations for the observed variation of
$R-K'$ with time.  First, it is possible that the color variations are
real.  Second, as suggested by Masetti et al (2000), it may be that
there are no true color variations, but that the afterglow exhibits
achromatic fluctuations on time scales short compared to the interval
between observations.  If so, the fluctuations need to have amplitude
$\sim 10\%$ and to occur on timescales $\delta t \la 3.3
\,\hbox{hours} = 0.04 t$ to explain the March 4.64 data, and $\delta t
\sim 7 \, \hbox{minutes} = 0.003 t$ to explain the March 3.21 data.
The latter in particular is physically implausible for an
afterglow expanding into a realistic external medium.  Finally, if the
error bars on the photometry are systematically underestimated, the
data may be reconciled with no color evolution.  In order to reduce
$\chi^2$ to $5$ (so that $\chi^2 / \hbox{dof} = 1$), we need to add a
systematic error contribution of $0.078$ magnitudes in quadrature to
all of the $K'$ and interpolated $R$ fluxes.  Given the discussion of
systematic errors above and our effort to adjust all R band fluxes to
the Henden et al calibration, this large an additional error is
unlikely.  We therefore believe that at least some of the measured
color variation is real.

\section{Spectral Energy Distributions} \label{sedsec}
The long spectral baseline between the K' band ($2.1 \micron$) and the
optical/UV bands allows one to obtain accurate spectral slope
measurements.  We have calculated the burst spectral
energy distribution at selected times based on the availability of
multiwavelength data.  Where necessary, flux measurements were
interpolated between adjacent data points at one wavelength in order
to determine a contemporaneous flux with another wavelength.  For this
operation, we always interpolated the light curves with good sampling
(R, K') to match the time of a sparsely sampled wavelength (UV $2825
\mbox{\AA}$, B).  Photometric zero points for the conversion of magnitudes to
flux density units were taken from Fukugita, Shimasaku, \& Ichikawa
(1995) for the optical, and from Campins, Rieke, \& Lebofsky (1985)
for the near-IR.

\begin{figure}
\plotone{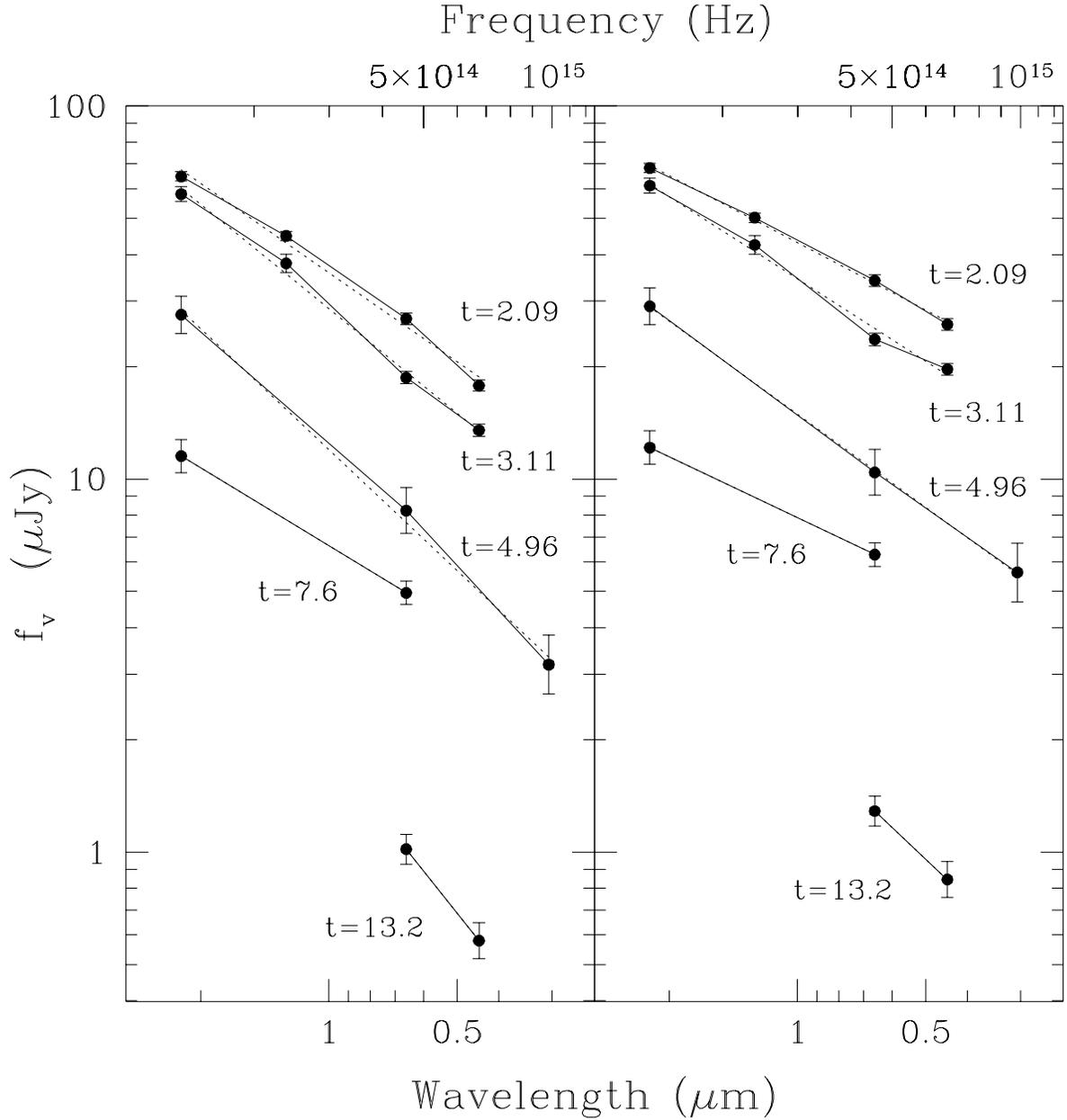}
\caption{
The spectral energy distribution of the GRB 000301c afterglow in the
observer-frame near-IR to near-UV is plotted for selected epochs.
Left panel: Spectral energy distributions (SEDs) corrected for Galactic
extinction only.  Note the slight but significant curvature of the SED
at the earliest epoch, when the errors are smallest.
Right panel: An additional correction for $A_V =
0.09$ magnitudes of SMC type dust
at an assumed host galaxy redshift $z=2.03$ has been
applied.  In both panels, the epochs of observation (from top to
bottom) are UT 2000 March 3.50, 4.52, 6.375, 9.00, and 14.60.  The
corresponding times since the GRB are labeled on the plot.
Solid lines connect flux densities sharing a common
epoch, and dotted lines show the best-fit single power law at each epoch.
\label{sedfig}}
\end{figure}

The resulting spectral energy distributions are shown in
figure~\ref{sedfig}.  The best-fit spectral slopes assuming unbroken
power law spectra are given in table~\ref{slopetable}, first
uncorrected for extinction; then with a correction for foreground
Galactic extinction ($E_{B-V} = 0.053$, Schlegel et al 1998) assuming
an $R_V \equiv A_V / E_{B-V} = 3.1$ extinction law; and finally with
an additional correction for possible extinction in the GRB host galaxy (see
below).  Also included are the slopes derived from R and K' data
alone, corrected only for Galactic extinction.

\begin{table}
\begin{tabular}{lclllll}
UT & $t$ & Filters        & Slope  & Slope  & Slope        & Slope \\
date & &                  & (none) & (MW)   & (MW $+$ host) & (R-K) \\
3.50 & 2.09    & B,R,J,K' & $-0.905\pm 0.026$ & $-0.789 \pm 0.026$ &
   $-0.596\pm 0.026$ & $-0.722 \pm 0.038$ \\
4.385 & 2.974  & B,R,J,K' & $-1.032\pm 0.032$ & $-0.911\pm 0.033$ &
    $-0.709\pm 0.032$ & $-0.901 \pm 0.052$ \\
4.52 & 3.11    & B,R,J,K' & $-1.049\pm 0.035$ & $-0.927\pm 0.036$ &
   $-0.727\pm 0.035$ & $-0.932 \pm 0.051$ \\
6.375 & 4.964  & UV,R,K'  & $-1.18\pm 0.10$  & $-1.03\pm 0.10$  &
   $-0.83\pm 0.10$ &  $-0.99 \pm 0.15$ \\
9.00  & 7.59   & R,K'     & $-0.78\pm 0.11$  & $-0.69\pm 0.10$  &
   $-0.54\pm 0.10$  &  $-0.69 \pm 0.10$ \\
14.60 & 13.19  & B,R      & $-1.64\pm 0.37$  & $-1.44\pm 0.37$  &
   $-1.07 \pm 0.36$  &  n/a
\end{tabular}
\caption{Spectral slopes $d \log{f_\nu} / d \log(\nu)$ 
of the GRB 000301c afterglow at selected epochs.
Column sub-headings indicate the type of extinction correction
applied, starting with none, then correcting for Milky Way extinction
only, and finally for both Milky Way extinction and host galaxy
extinction.  The last column gives the slope measured using the R and
K' filters alone, corrected for Milky Way extinction only,
to isolate the time evolution of the burst from possible
systematic differences among filters.  Host galaxy extinction
corrections assume $A_V = 0.09$ at $z=2.03$ with a Small Magellanic
Cloud extinction law (Pei 1992).
\label{slopetable}
}
\end{table}

The broad band spectral energy distributions plotted in
figure~\ref{sedfig} show no compelling evidence for extinction at the
redshift of the GRB.  (This is a change from the first preprint of
this paper, largely due to revised calibration of the March 6.375 UV
data point at $3050 \mbox{\AA}$.)  However, a modest amount of host galaxy
extinction remains consistent with the data.

The $2175$\AA\ dust absorption feature falls into the observed R band
at the redshift $z=2.03$ of this burst.  Assuming that there is no
intrinsic emission feature at this wavelength in the GRB afterglow
spectrum, we can place an upper limit of $A_V \la 0.1$ magnitude
for Milky Way type dust at $z=2.03$.  Dust
with a different reddening law is less strongly constrained.  A Small
Magellanic Cloud reddening law is plausible for $A_V \la 0.12$, while
a Large Magellanic Cloud (LMC) law is plausible for $A_V \la 0.2$.
The measure of plausibility here is that a single power law should
approximately fit the spectrum at any given epoch.  As an illustrative
case, we have shown in figure~\ref{sedfig} the spectral energy
distributions corrected for $A_V = 0.09$ of SMC type dust.  This
correction gives the minimum $\chi^2 / \dof$ ($1.45$ for 4 degrees of freedom)
for the residuals of the plotted spectral energy distributions
relative to the spectral slopes reported in table~\ref{slopetable}.
LMC extinction does marginally worse than SMC extinction, while any
amount of Milky Way extinction in the host degrades $\chi^2$.  We have
used the analytic extinction law fitting forms of Pei (1992) in
deriving these estimates.  Jensen et al (2000) have applied a similar
analysis incorporating optical spectra as well as broadband colors.
They also find a significantly better fit for SMC extinction than for
either MW or LMC extinction, and derive $A_V = 0.14 \pm 0.01$
magnitudes for the SMC model.  Using our multiple epoch SEDs, we find
a somewhat worse $\chi^2 / \dof = 1.98$ for their extinction estimate than for
ours, but the two results are probably consistent within the errors,
especially if there are unidentified systematic errors of $\ga 0.08$
magnitude in the photometry (see section~\ref{colsec}).

The apparent weakness of the $2175$\AA\ feature in the extinction curve
of the host galaxy is reminiscent of dust attenuation laws for
actively star forming galaxies (e.g., the Magellanic Clouds [Pei 1992
and references therein] and starburst galaxies [Gordon, Calzetti, \&
Witt 1997]).  This may be further circumstantial
evidence linking GRBs to actively star forming galaxies.
Alternatively, such an extinction law might be observed if GRBs
preferentially destroy the small carbonaceous particles thought to be
carriers of the $2175$ \AA\ feature, but this explanation would only
work if much of the dust optical depth arises near the maximum radius
where the burst can destroy grains (cf.\ Waxman \& Draine 2000;
Fruchter, Krolik, \& Rhoads 2000).

% Note for following: Bertoldi point moved to 2.1 \pm 0.3 mJy
% in the Berger et al manuscript.

In order to determine physical parameters of the afterglow, we need to
measure the peak flux density and the locations of breaks in the
afterglow spectrum (Wijers \& Galama 1999).  We now do this (insofar
as possible) by combining our optical-IR spectral slope measurements
with the submillimeter and radio data.  We are looking for four numbers:
The frequency $\nu_{hbox{max}}$ and flux density $f_{\nu,\hbox{max}}$
at the peak in $f_\nu$; the cooling frequency $\nu_c$, and the
self-absorption frequency $\nu_{\hbox{abs}}$.  The spectral slope is
expected to be $-p/2$ for $\nu > \nu_c$, $-(p-1)/2$ for
$\nu_{\hbox{max}} < \nu < \nu_c$, $+1/3$ for $\nu_{\hbox{abs}} < \nu <
\nu_{\hbox{max}}$, and $+2$ for $\nu < \nu_{\hbox{abs}}$ (Sari, Piran,
\& Narayan 1998).  Here $p$ is the power law index of electrons
recently accelerated at the external shock of the expanding GRB
remnant.

\begin{figure}
\plotone{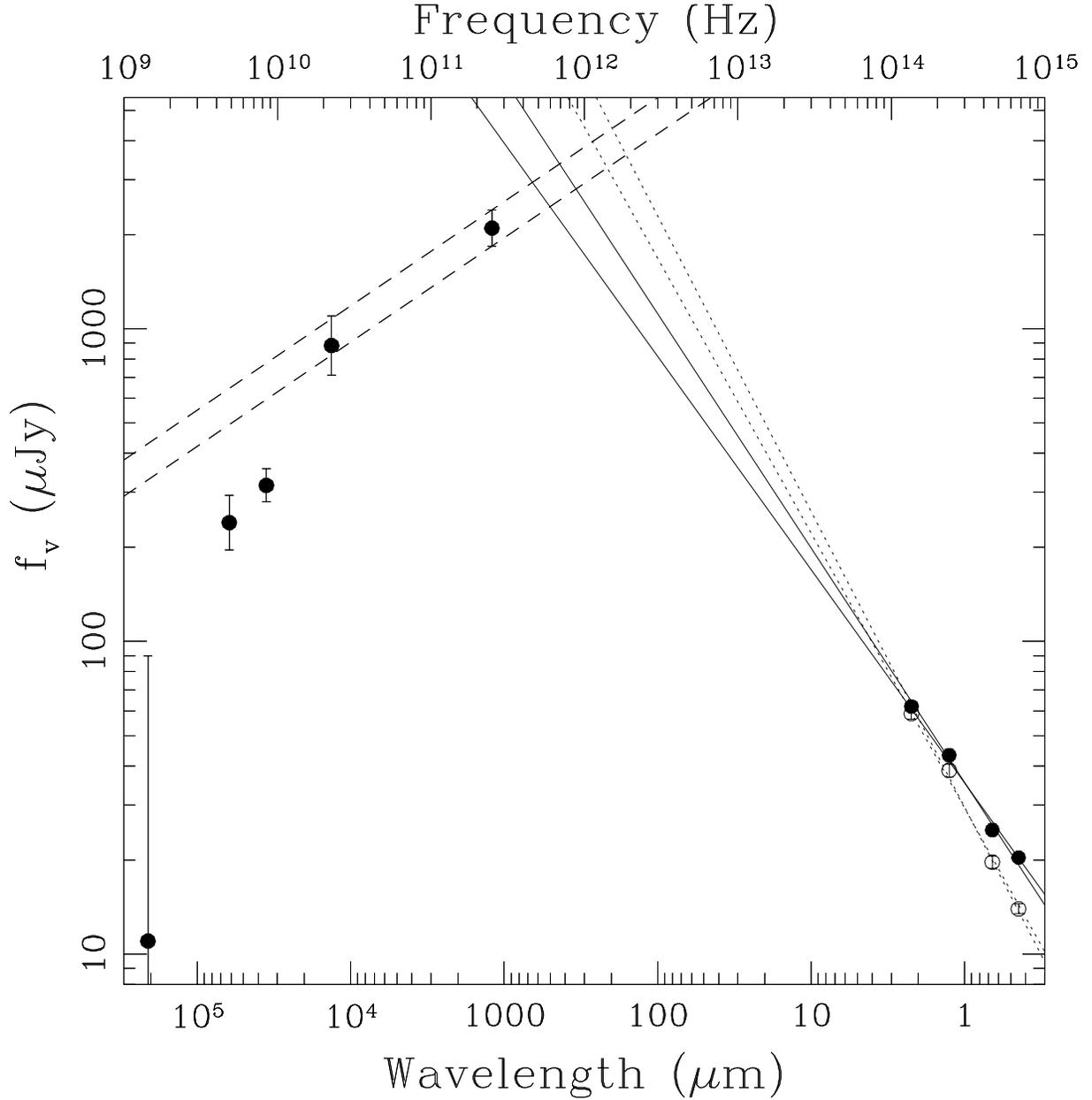}
\caption{ The spectral energy distribution of the GRB 000301c
afterglow from $250 \GHz$ to $0.44 \micron$ on UT 2000 March 4.385
(2.974 days post-GRB), and at $1.43$, $4.86$, $8.46$, and $22.5 \GHz$
on UT 2000 March 5.67 (4.26 days post-GRB).  Filled points show the
photometric data corrected for both Galactic foreground and GRB host
galaxy extinction (see text). Solid lines show the $\pm 1 \sigma$
fitted power law slopes through these fully corrected optical/IR data.
Open points show the optical/IR data corrected for only foreground
Galactic extinction, and dotted lines show the $\pm 1 \sigma$ fitted
power law slopes through these data points.  Finally, dashed lines
show the $f_\nu \propto \nu^{1/3}$ slope expected between the
self-absorption frequency and the peak in $f_\nu$.
\label{sedwide}}
\end{figure}

Extrapolating the optical-IR spectral slopes to lower frequencies, we
see that a strong spectral break is required near or above the $250
\GHz$ measurement by Bertoldi (2000) on March 4.385.  The radio data
from March 5.67 (Berger et al 2000) are compatible with $f_\nu
\propto \nu^{1/3}$ between $22$ and $250 \GHz$, so we determine
$\nu_{\hbox{max}}$ and $f_{\nu,\hbox{max}}$ by extrapolating this
behavior until it intersects the extrapolation from optical-IR data
(see figure~\ref{sedwide}).  Using fluxes corrected for both Galactic dust
and $A_V = 0.09$ magnitude of SMC type extinction at $z=2.03$, we
obtain $\log(\nu_{\hbox{max}}/\Hz) = 11.81 \pm 0.10$ and
$\log(f_{\nu,\hbox{max}} / \mJy ) = 0.46 \pm 0.05$, where the error
bars account only for photometric errors on the data.  If we do not
apply any correction for host galaxy extinction, we instead obtain
$\log(\nu_{\hbox{max}}/\Hz) = 12.16 \pm 0.08$ and
$\log(f_{\nu,\hbox{max}} / \mJy ) = 0.58 \pm 0.05$.  We can also
estimate $\nu_{\hbox{abs}}$ using the radio data from Berger
et al (2000) that is shown in figure~\ref{sedwide}.  
We conservatively estimate $3 \GHz \la
\nu_{\hbox{abs}} \la 15 \GHz$ at March 5.66.  Frail (personal
communication) reports a best fit value of about $6.8 \GHz$.

% The spectral slope at radio to millimeter wavelengths is generally
% expected to be $+1/3$ at these early times.  This slope arises from
% the low frequency tail of electron synchrotron emission at frequencies
% below the spectral peak of the least energetic (and most numerous)
% relativistic electrons.  It should continue from the $f_\nu$ peak down
% to the self-absorption frequency $\nu_{\hbox{abs}}$, below which we
% expect $f_\nu \propto \nu^2$ (Sari, Piran, \& Narayan 1998).

The location of the cooling frequency $\nu_c$ is harder to constrain,
not least because it is a relatively modest break of $0.5$ in spectral
index and because this burst is not very well observed at X-ray
wavelengths.  If we take the afterglow behavior to be reasonably
described by the ``standard'' model at the earliest observed times,
then the expected behaviors are $f_\nu \propto \nu^{-(p-1)/2}$ and
$f_\nu \propto t^{-3 (p-1) / 4}$ for $\nu < \nu_c$, or $f_\nu \propto
\nu^{-p/2}$ and $f_\nu \propto t^{1/2 - 3 p / 4}$ for $\nu > \nu+c$.
While this model is based on a uniform ambient medium and spherically
symmetric burst, it is also valid for GRBs collimated into an angle
$\zeta$ at early times, while $\Gamma > 1/\zeta$ (Rhoads 1997, 1999).
The early time R band data (excluding the discrepant March 2.906 data
point)  gives $f_\nu \propto t^{-0.71}$.  The spectral slope is $f_\nu
\propto \nu^{-0.9}$ if we omit host galaxy extinction corrections, but
could be as blue as $\nu^{-0.6}$ for moderate host galaxy extinction.
Both theoretical models of electron acceleration in relativistic
shocks (Bednarz \& Ostrowski 1998; Gallant, Achterberg, \& Kirk 1999) and
experience with other afterglows give typical values $p \approx 2.3$.
Comparable values of $p$ are marginally consistent with the early time
data if we adopt $A_V \ga 0.12$ in
the host galaxy and $\nu_c > \nu \approx 10^{15} \Hz$ at $t = 3$ days.
If we require $p \ga 2$, then $\nu_c < 10^{15} \Hz$ yields much poorer
agreement.  Only if we allow $1.4 \la p \la 1.6$ can we find a viable model
having $\nu_c$ between optical and radio wavelengths.  
However, a new parameter (the upper cutoff in the electron energy
spectrum) must be added to the ``standard'' model to accomodate $p<2$
and still keep a finite energy in relativistic electrons.
Additionally, it is very difficult to fit the late time decay $f_\nu
\propto t^{-2.8}$ with any $p<2$ model; the only viable possibility we
presently know of is the ``naked afterglow'' model of Kumar \&
Panaitescu (2000b).  If we consider the early time K' light curve,
$f_\nu \propto t^{-0.1}$, and no regime under the standard model offers a
self-consistent estimate of $p$.
Overall, we regard $p \ga 2$ and $\nu_c \ga
10^{15} \Hz$ around UT 2000 March 3.5 as the most plausible solution.

\section{Physical Parameters of the Afterglow}
We can use the observed spectral energy distribution of GRB 000301C to
place interesting limits on some of the afterglow's physical
parameters, using the method of Wijers \& Galama (1999).  To do so, we
assume that the afterglow is reasonably approximated by a spherical
burst (or a section thereof) expanding into a uniform ambient medium
up to the time of our SED measurement.  Even
if the light curve break is attributed to a jet, the method should
work for times before the break.  We take the
spectrum to peak at either of the values derived above
($\log(\nu_{\hbox{max}}/\Hz) = 11.81$ and $f_{\nu,\hbox{max}} = 2.9
\mJy$ with the host galaxy extinction correction, or
$\log(\nu_{\hbox{max}}/\Hz) = 12.16$ and $f_{\nu,\hbox{max}} =3.8
\mJy$ with only Milky Way extinction corrections).
In addition, we take $\log(\nu_c / \Hz) \ge 15$.  Using
$\nu_{\hbox{abs}} = 6.8 \GHz$ then gives ``best estimate'' limits of
$E > 3\times 10^{53} \erg \times \Omega / (4 \pi)$,
$\xi_e \ge 0.11$,
$\xi_B \le 1.6 \times 10^{-3}$,
and $n \ga 1 cm^{-3}$.
Here $E$ is the kinetic energy of the ejecta, $\xi_e$ and $\xi_B$ are
the fractions of the local energy that go into relativistic electrons
and magnetic fields immediately behind the expanding GRB remnant blast
wave, and $n$ is the number density of the ambient medium.  We obtain
limits rather than measurements because of the relatively weak
constraints on $\nu_c$ and $\nu_{\hbox{abs}}$.  We have taken the most
conservative pair of ($\nu_{\hbox{max}}$, $f_{\nu,\hbox{max}}$) in
deriving these limits on physical quantities.  The physical quantities
scale with $\nu_{\hbox{abs}}$ as $E \propto \nu_{\hbox{abs}}^{-5/6}$,
$\xi_e \propto \nu_{\hbox{abs}}^{5/6}$, $\xi_B \propto
\nu_{\hbox{abs}}^{-5/2}$, and $n \propto \nu_{\hbox{abs}}^{25/6}$
(Wijers \& Galama 1999).  
If we allow $3 \GHz \la \nu_{\hbox{abs}} \la 15 \GHz$, and further take
the one-sigma variations on $\nu_{\hbox{max}}$ and $f_{\nu,\hbox{max}}$
to weaken the limits, we obtain more conservative limits of
$E > 1.3\times 10^{53} \erg \times \Omega / (4 \pi)$,
$\xi_e \ge 0.04$,
$\xi_B \le 0.02$,
and $n \ga 0.014 cm^{-3}$.
We have used $p=2.3$, but other values affect the results only weakly
except for $\xi_e$, whose behavior approaches $(p-2)^{-1}$ as $p
\rightarrow 2$.
If the ill-constrained cooling frequency is substantially above
$10^{15} \Hz$ at 3 days, then the bounds on $E$ and $\xi_e$ rise as
$\nu_c^{1/4}$, $\xi_B$ falls as $\nu_c^{-5/4}$, and $n$ rises as
$\nu_c^{3/4}$ (Wijers \& Galama 1999).  On the other hand, $\nu_c
\ll 10^{15} \Hz$ requires $p \sim 1.5$.  In this case the equations
describing the afterglow and the inversion of measured quantities to
obtain physical parameters would require new formulae.

It is interesting to compare our estimate of $E$ with the gamma ray
fluence of the burst.  Jensen et al (2000) estimate a fluence of $2.1
\times 10^{-6} \erg \cm^{-2}$ in the $25$ to $100 \keV$ band and a
corresponding energy of $2.3 \times 10^{52} \erg \times \Omega / (4
\pi)$.  They also report a similar fluence in the $150$ to $1000 \keV$
band, so the total energy might be $\sim 5 \times 10^{52} \erg \times
\Omega / (4 \pi)$.  Thus, our most conservative limits $E > 1.3\times
10^{53} \erg \times \Omega / (4 \pi)$ implies that $\la 0.4$ of the
blast wave energy was emitted in gamma rays, while for our best
estimate $E > 3\times 10^{53} \erg \times \Omega / (4 \pi)$, this
fraction is reduced to $\la 0.15$.

\section{Discussion}
GRB 000301c is the third burst for which a strong break in the light
curve is clearly observed.  Several classes of breaks are predicted by
fireball models.  The most basic of these are due to features in the
synchrotron spectrum moving through the observed bandpass (e.g.,
Paczy\'{n}ski \& Rhoads 1993; Sari, Piran, \& Narayan 1998).  However,
this class of features predicts relatively modest changes in light
curve slope, with the break occurring first at short wavelengths and
evolving to longer ones.  Jetlike burst ejecta, on the other hand, are
expected to give strong breaks that are essentially independent of
wavelength (Rhoads 1997, 1999; Sari, Piran, \& Halpern 1999),
and the observed breaks have generally been
interpreted as evidence for collimation of the GRB ejecta (e.g., in
GRB 990123, by Castro-Tirado et al 1999, Kulkarni et al 1999, Fruchter
et al 1999, and Galama et al 1999; and in GRB 990510, by Stanek et al
1999 and Harrison et al 1999).  A difficulty with this model is that
the predicted break is quite gradual (Rhoads 1999; Panaitescu \&
Meszaros 1999; Moderski, Sikora, \& Bulik 2000; Kumar \& Panaitescu
2000a), while observed breaks are rather sharp.

The current burst is no exception.  The prediction of Rhoads (1999)
for the light curve around the break time for a collimated jet is
\begin{equation}
f_\nu \propto {  [t/t_b]^{-3(p-1)/4} \over 
\left(1 + 3.72 \, [t/t_b]^{2/5}\right)^{5/2} * \left(1 + 2.07 \,
[t/t_b]^{5/12}\right)^{3(p-1)/5} }
\end{equation}
where $t_b$ is the fiducial break time defined in Rhoads (1999).
The break in this predicted light curve is extremely broad, and would give
$\chi^2$ little better than a single power law in fitting the observed
break in either $K'$ or $R$ band.  The model curve is
based on numerical integration of the remnant's dynamical equations,
and ignores differences in light travel time between the center and
edge of the remnant, which will only smooth the break further (e.g.,
Moderski et al 2000; Panaitescu \& Meszaros 1999).

If we ignore the issue of break sharpness and fit a collimated jet
model to the observed R band light curve, we can infer the opening
angle from the measured break time.  To do so, we need a reasonable
measurement of $t_b$ and crude estimates of $\Omega/E$ and $n$, since the
inferred opening angle scales as $(t_b^3 n \Omega/E )^{1/8}$ (Rhoads 1999).
We use $t_b = 5.1$ days and $E = 3 \times 10^{53} \erg \times \Omega / (4
\pi)$.  To estimate $n$ more precisely, we use the column
density $N(HI) \approx 10^{21.2 \pm 0.5}$ inferred from Lyman $\alpha$
absorption (Jensen et al 2000) and estimate the linear size of the
source as $\la 0.2''$ based on its nondetection in late HST images.
This implies a number density $n \ga 0.4$, consistent with our earlier
estimate. Using $n \approx 1$, the inferred opening angle becomes
$2.5^\circ$, or $10^{-3}$ of the sky if the jet is bipolar.

The transition to the nonrelativistic regime has been proposed as
another mechanism for light curve breaks both in this burst (Dai \& Lu
2000) and others (Dai \& Lu 1999).  However, we do not know of a
detailed calculation of the sharpness of this break, making a fair
evaluation of this possibility difficult.  Light travel time effects
seem likely to broaden this feature to $\delta t / t \sim 1$, as with
most other features.

A final possible cause for sharp breaks in GRB afterglow light curves
is discontinuities in the ambient density distribution.  Assuming that
the density is a function of radius alone, a minimum timescale for breaks
due to such discontinuities is $\Delta t \ga t$, where $t$ is the time
elapsed in the observer's frame since the burst and $\Delta t$ the
characteristic duration of a light curve feature.  This duration is set by
differential light travel time effects between material moving along
the line of sight and off-axis material moving in direction
$1/\Gamma$, and is a rough minimum for any afterglow light curve
feature provided the ambient medium density is approximately independent of
angle from the line of sight.  The time required for material already
in the expanding blast wave to cool by adiabatic losses is also
relevant for determining the sharpness of a density discontinuity
break, as is the emission from angles $> 1/\Gamma$ off the line of
sight (Kumar \& Panaitescu 2000b).

If we believe that the observed $R-K'$ variations are real,
then the greatest difficulty posed by the observations of GRB
000301c is in finding a model whose light curve steepens in $K'$ {\it
before} it steepens in R.  For most mechanisms, breaks will occur
either first at short wavelengths (e.g. the cooling break), or
simultaneously at all wavelengths (e.g. ``beaming'' breaks).  One
speculative way out is to suppose that a discontinuity in the ambient
density is encountered while the cooling break is between the R and K'
filters.  The predicted appearance of an afterglow at frequencies above
and below this break is expected to differ qualitatively: A high
frequency image would show an annular structure and a low frequency
image a more nearly filled disk.  This is caused by the difference in
the apparent dynamical age of the remnant along the line of sight
(where we see things changing quickly) and near the edge of the
observed afterglow (where light travel time is larger, and we see
material at an earlier and hotter stage of its evolution).  (E.g.,
Granot, Piran, \& Sari 1999.)  Now, the same variation of ``lookback
time'' from the center to edge of the afterglow implies that we see
the effect of an ambient density drop first in the middle of the
afterglow, and that the fractional effect of such a discontinuity on
the afterglow flux will initially be larger at long wavelengths than
short ones.  This mechanism can reproduce the sign of the observed
effect.  Detailed calculations would be necessary to see if it can
approach the observed magnitude, given that the two filters are separated
by only a factor of $3$ in wavelength.

\section{Summary}

We present a K' band ($2.1 \micron$) light curve of the GRB 000301c
afterglow, combining four epochs from a Target of Opportunity program
at the NASA Infrared Telescope Facility with two additional
measurements from Calar Alto (Stecklum et al 2000) and Subaru
(Kobayashi et al).  This light curve can be well fitted by a broken
power law evolution, with a very flat early time slope ($t^{-0.1}$), a
steep late time slope ($t^{-2.3}$), and a rather sharp break.  A
similar fit for R band ($0.7 \micron$) data from the literature yields
steeper slopes ($-0.7$ and $-2.8$) and a later break time.  A
detailed analysis of the $R-K'$ colors shows modest deviations from
constant color, which are significant at the $\sim 2 \sigma$ level.
These may indicate the presence of systematic errors $\ga 0.08$
magnitude due to inhomogeneous data sets.  However, we believe that
the likely level of such systematic errors is $\la 0.05$ magnitude,
and that at least part of the color variability is likely real.

The strong break and steep late time slope in the light curves are
reminiscent of GRB 990510, which has been interpreted as a collimated
burst (Stanek et al 1999; Harrison et al 1999; Kumar \& Panaitescu
2000a).  If we fit a collimated burst model to the R band light curve,
the estimated opening angle becomes $2.5^\circ$.

However, while a jet model can reproduce the size of the break, it
does not provide a natural explanation for the observed rapidity of
the break.  Nor, however, do other models. Indeed, there is
considerable variability in the R band light curve of GRB 000301C on
time scales $\delta t / t \le 0.4$.  This highlights the potential for
unusual temporal variability in GRB afterglows.

Fitting a standard synchrotron spectral energy distribution to the
burst, we place the peak of $f_\nu$ around $3.4 \mJy$ at about $
10^{12} \Hz$ ($300 \micron$).  The dominant uncertainty in this
measurement is the correction for extinction in the GRB host galaxy.
Random errors are $\approx 0.10$ dex in $\nu_{\hbox{max}}$ and $0.05$
dex in $f_{\nu,\hbox{max}}$, while the correction for host galaxy
extinction is uncertain at perhaps twice this level.  These strong
constraints on the $f_\nu$ peak are possible because of the precise
optical-IR spectral slope measurements afforded by K' observations.

Combining the measured spectral peak with constraints on the cooling
frequency and self-absorption frequency, we infer that the blast wave
energy required to power this afterglow was $E > 3 \times 10^{53}
\erg$ if isotropic.  The corresponding efficiency of gamma ray
production in the burst was $\la 0.15$.

\acknowledgements
It is a pleasure to thank the NASA IRTF director and staff for making
this observing program possible.  Special thanks are due to Bob
Joseph, for arranging the target of opportunity mechanisms; Bill
Vacca, for help with observing strategies; and Bill Golisch, Paul
Fukumura-Sawada, and Dave Griep for observing.  We also thank Sylvio Klose,
Dale Frail, and Naoto Kobayashi for useful communications.
JER's work is supported by an Institute Fellowship at STScI.

\end{document}